\documentclass[pre,showpacs,superscriptaddress]{revtex4}
\usepackage{graphicx}
\usepackage{epsfig}
\begin{document}
\draft
\newcommand{\be}{\begin{equation}}
\newcommand{\ee}{\end{equation}}
\newcommand{\bea}{\begin{eqnarray}}
\newcommand{\eea}{\end{eqnarray}}
\newcommand{\mbf}{\mathbf}

\title{Mitigating the effects of measurement noise on Granger causality}
\author{Hariharan Nalatore}
\email{hnalatore@bme.ufl.edu} \affiliation{The J. Crayton Pruitt
Family Department of Biomedical Engineering, University of Florida,
Gainesville, FL 32611, USA}
\author{Govindan Rangarajan}
\email{rangaraj@math.iisc.ernet.in} \affiliation{ Department of
Mathematics, Indian Institute of Science,  Bangalore - 560 012,
India}
\author{Mingzhou Ding}
\email{mding@bme.ufl.edu} \affiliation{The J. Crayton Pruitt Family
Department of Biomedical Engineering, University of Florida,
Gainesville, FL 32611, USA}


\begin{abstract}

Computing Granger causal relations among bivariate experimentally
observed time series has received increasing attention over the past
few years. Such causal relations, if correctly estimated, can yield
significant insights into the dynamical organization of the system
being investigated. Since experimental measurements are inevitably
contaminated by noise, it is thus important to understand the
effects of such noise on Granger causality estimation. The first
goal of this paper is to provide an analytical and numerical
analysis of this problem. Specifically, we show that, due to noise
contamination, (1) spurious causality between two measured variables
can arise and (2) true causality can be suppressed. The second goal
of the paper is to provide a denoising strategy to mitigate this
problem. Specifically, we propose a denoising algorithm based on the
combined use of the Kalman filter theory and the
Expectation-Maximization (EM) algorithm. Numerical examples are used
to demonstrate the effectiveness of the denoising approach.

\end{abstract}


\maketitle

\section{Introduction}

Granger causality \cite{Granger} has become the method of choice to
determine whether and how two time series exert causal influences on
each other. In this method one starts by modeling simultaneously
acquired time series as coming from a multivariate or vector autoregressive
(VAR) stochastic process. One time series is said to have a causal
influence on the other if the residual error in the autoregressive
model of the second time series (at a given point of time) is
reduced by incorporating past measurements from the first. This
method and related methods have found applications in a wide variety of fields including
physics \cite{Blinowska,Marinazzo,Verdes,Rosenblum,Hu,Xu},
economics \cite{Granger,Thornton,Hall,Hiemstra} and neuroscience
\cite{Ding,Ding2}. Its nonlinear extension has recently appeared in
\cite{Chen} and has been applied to study problems in condensed
matter physics \cite{Rajesh}.

The statistical basis of Granger causality estimation is linear
regression. It is known that regression analysis is sensitive to the
impact of measurement noise \cite{wayne}. Given the inevitable
occurrence of such noise in experimental time series, it is
imperative that we determine whether and how such added noise can
adversely affect Granger causality estimation. Previous studies
\cite{Newbold} have suggested that such adverse effects can indeed
occur. In this paper, we make further progress by obtaining
analytical expressions that explicitly demonstrate how the interplay
between measurement noise and system parameters affects Granger
causality estimation. Moreover, we show how this deleterious effect
of noise can be reduced by a denoising method, which is based on the
Kalman filter theory and the Expectation-Maximization (EM)
algorithm. We refer to our denoising algorithm as the KEM (Kalman EM) denoising
algorithm.

The organization of this paper is as follows. In Section 2, we start
by introducing an alternative formulation of Granger causality
\cite{Pierce} and proceed to outline a framework within which the
effects of added (measurement) noise on the estimation of
directional influences in bivariate autoregressive processes can be
addressed. To simplify matters, we then consider a bivariate first
order autoregressive (AR(1)) process in Section 3. Here explicit
expressions for the effect of noise on Granger causality are
derived. These expressions allow us to show that, for two time
series that are unidirectionally coupled, spurious causality can
arise when noise is added to the driving time series and true
causality can be suppressed by the presence of noise in either time
series. The theoretical results are illustrated by numerical
simulations. In Section 4, we briefly introduce the KEM denoising
algorithm and apply it to the example considered in Section 3. Our
results show that the KEM algorithm can mitigate the effects of
noise and restore the true causal relations between the two time
series. In section 5, we consider a coupled neuron model which
produces time series that closely resemble that recorded in neural
systems. The effect of noise on Granger causality and the
effectiveness of the KEM algorithm in mitigating the noise effect
are illustrated numerically. Our conclusions are given in Section 6.

\section{Theoretical Framework}

Consider two time series $X(t)$ and $Y(t)$. To compute Granger
causality, we model them as a combined bivariate autoregressive
process of order $p$. In what follows, the model order $p$ is
assumed to be known, since this aspect is not central to our
analysis. The bivariate autoregressive model can then be represented
as:
\begin{eqnarray}\label{bivariate_AR}
\sum_{k=0}^p [a_k X(t-k)+b_k Y(t-k)] = E_1(t),\\
\sum_{k=0}^p [c_k X(t-k) +d_k Y(t-k)] = E_2(t),
\end{eqnarray}
where $a_k$, $b_k$, $c_k$, and
$d_k$ are the AR coefficients and $E_i(t)$ are the
temporally uncorrelated residual errors.

For our purposes, it is more convenient to rewrite the above
bivariate process as two univariate processes (this can always be
done according to \cite{Pierce}): \be
 P_{1}(B)X(t)= {\xi(t)}; \ \ \
 P_{2}(B)Y(t)= {\eta(t)},
\label{Eq1}\ee
where $B$ is the lag operator defined as $B^k
X(t)=X(t-k)$ and $P_1$ and $P_2$ are polynomials (of possibly
infinite order) in the lag operator $B$. It should be noted that the
new noise terms $\xi(t)$ and $\eta(t)$ are no longer uncorrelated.
Let $\gamma_{12}(k)$ denote the covariance at lag $k$ between these
two noises. \be
 {\gamma_{12}(k)}\equiv {\rm cov} ( {\xi(t))},  {\eta(t-k)})~~~~~~~
 {k = ..., -1, 0, 1 ...}\ .
\ee
A theorem by Pierce and Haugh \cite{Pierce} states that $ {Y(t)}$
causes $ {X(t)}$ in Granger sense if and only if
\be
 {\gamma_{12}(k)\neq 0}~ \rm{for~ some}~ k > 0.
\ee
Similarly $ {X(t)}$ causes $ {Y(t)}$ if and only if
$ {\gamma_{12}(k)\neq 0}$ \rm{for some} $k < 0$.

Now we add measurement noises $\xi'(t)$ and $\eta'(t)$ to $X(t)$ and $Y(t)$ respectively:
\begin{eqnarray}
 {X^{(c)}(t)}=  {X(t)}+ {\xi'(t)},\\
 {Y^{(c)}(t)}=  {Y(t)}+ {\eta'(t)}.
\end{eqnarray}
Here $ {\xi'(t)}$, $ {\eta'(t)}$ are
uncorrelated white noises that are uncorrelated with
$ {X(t)}, {Y(t)},  {\xi(t)}$ and $ {\eta(t)}.$ Following Newbold \cite{Newbold},
the above equations can be rewritten as
\begin{eqnarray}
P_{1}(B) {X}^{(c)}(t)= P_{1}(B) {X}(t)+P_{1}(B) {\xi}'(t),\\
P_{2}(B) {Y}^{(c)}(t)=
P_{2}(B) {Y}(t)+P_{2}(B) {\eta}'(t).
\end{eqnarray}
Using Eq. (\ref{Eq1}) we get
\begin{eqnarray}
P_{1}(B) {X}^{(c)}(t)=  {\xi}(t)+P_{1}(B) {\xi}'(t),\\
P_{2}(B) {Y}^{(c)}(t)=
 {\eta}(t)+P_{2}(B) {\eta}'(t).
\end{eqnarray}

Following the procedure in Granger and Morris \cite{Morris}, the linear
combination of white noise processes on the right hand
sides can be rewritten in terms of invertible moving average
processes \cite{Maravall}:
\begin{eqnarray}
 {\xi}(t)+P_{1}(B) {\xi}'(t)=P_{3}(B) {\xi}^{(c)}(t),\\
 {\eta}(t)+P_{2}(B) {\eta}'(t)=P_{4}(B) {\eta}^{(c)}(t),
\end{eqnarray}
where $ {\xi^{(c)}}$ and $ {\eta^{(c)}}$ are again
uncorrelated white noise processes.  Thus
we get
\begin{eqnarray}
P_{3}^{-1}(B)P_{1}(B) {X}^{(c)}(t)=  {\xi^{(c)}}(t), \cr
P_{4}^{-1}(B)P_{2}(B) {Y}^{(c)}(t)=  {\eta^{(c)}}(t).
\end{eqnarray}
This is again in the form of two univariate AR processes. Therefore
the theorem of Pierce and Haugh can be applied to yield the result
that the noisy signal $ {Y^{(c)}(t)}$ causes $ {X^{(c)}(t)}$ in
Granger sense if and only if
\be {\gamma_{12}^{(c)}(k)} \equiv {\rm
cov}( {\xi}^{(c)}(t),  {\eta}^{(c)}(t-k))\neq 0, \ee for some $ {k}
> 0.$ Similarly $ {X^{(c)}(t)}$ cause $ {Y^{(c)}(t)}$ if and only if
\be \gamma_{12}^{(c)}(k)\neq 0, \ee for some $ {k < 0}$.

We can relate $  \gamma_{12}^{(c)}$ to $  \gamma_{12}$ as follows.
Consider the corresponding covariance generating functions (which
are nothing but the $z$-transforms of the cross-covariances)
\bea\label{covgenfn}
 \bar{\gamma}_{12}(z) &= &\sum_{k=-\infty}^{\infty} \gamma_{12}(k)z^{k},\cr
 \bar{\gamma}_{12}^{(c)}(z) &= &\sum_{k=-\infty}^{\infty}\gamma_{12}^{(c)}(k)z^{k}.
\eea
We can show that \cite{Newbold}
\be
\bar{\gamma}_{12}^{(c)}(z)=P_{3}^{-1}(z)P_{4}^{-1}(z^{-1})\bar{\gamma}_{12}(z).
\label{Eq2}\ee
Even if $ {\gamma_{12}(k)= 0}$ for all $k < 0$
(i.e. $ {X}$ does not cause  $ {Y}$) it is possible
that $ \gamma_{12}^{(c)}(k)\neq 0$ for some negative $k$ because of
the additional term $P_{3}^{-1}(z)P_{4}^{-1}(z^{-1})$ that has
been introduced by the measurement noise.
This gives rise to the spurious Granger causality, ($ {X}^{(c)}$ causes
$ {Y^{(c)}}$), which is a consequence of the added measurement noise.

\section{A Bivariate AR(1) Process}

In the previous section, we demonstrated that measurement noise can
affect Granger causality. But the treatment given was quite general
in nature. In this section we specialize to a simple bivariate AR(1)
process and obtain explicit expressions for the effect of noise on
Granger causality.

Consider the following bivariate AR(1) process
\begin{eqnarray}
 {X(t)}&=& a  {X(t - 1)} + b  {Y (t - 1)} +  E_1(t), \nonumber \\
 {Y(t)}&=& d  {Y(t-1)}+ E_2(t).
\label{Eq3}\end{eqnarray} From the above expressions, it is clear
that $Y$ drives $X$ for nonzero values of $b$ and $X$ does not drive
$Y$ in this model. More specifically, we see that $Y$ at an earlier
time $t-1$ affects $X$ at the current time $t$. There is no such
corresponding influence of $X$ on $Y$.

When noises $\xi'(t)$ and $\eta'(t)$ with variances
$\sigma_{\xi'}^{2}$ and $\sigma_{\eta'}^{2}$, respectively, are
added to the data generated by Eq. (\ref{Eq3}), after some algebra
(see Appendix for details),
we find the following expressions for $P_3(B)$ and $P_4(B)$:
\be \label{P3P4}
P_3(B)=1+a_1'B+a_2' B^2; \ \ \ P_4(B)=1-d^{\ '} B. \ee Here \be d^{\
'} = \frac{s{\pm{\sqrt{s^{2}-{4}}}}}{2},
\ee where
\be
s\equiv(\frac{1}{d}+d)+\frac{1}{d}\frac{ {\sigma_{\eta}^{2}}}{
{\sigma_{\eta'}^{2}}}. \ee The expressions for $a_1'$ and $a_2'$ are
very long and for our purposes it is sufficient to note that they go
to zero as the added noise goes to zero (as expected). We see that $
|s| > 2 $ for any value of $d$,
 ${\sigma_{\eta}^{2}}$ and ${\sigma_{\eta'}^{2}}$.
Therefore $\sqrt{S^{2}-{4}}$ and hence $d{\ '}$ are well defined.
We also have the following results:
\begin{description}
\item{a)} As $|d|\rightarrow 0$,
$|d^{\ '}|<|d|\rightarrow 0$;
\item{b)} As $d\rightarrow 1$, $d^{\ '}\rightarrow 1 +
\frac{ {\sigma_{\eta}^{2}}}{2 {\sigma_{\eta'}^{2}}}
-\frac{ {\sigma_{\eta}^{2}}}{2 {\sigma_{\eta'}^{2}}}\sqrt{1+4/\frac{ {\sigma_{\eta}^{2}}}{ {\sigma_{\eta'}^{2}}}}$;
\item{c)} As the ratio
$\frac{ {\sigma_{\eta}^2}}{ {\sigma_{\eta'}^2}}\rightarrow
0,  d^{\ '}\rightarrow d$;
\item{d)}  As the ratio $\frac{ {\sigma_{\eta}^2}}{ {\sigma_{\eta'}^2}}\rightarrow
\infty, d^{\ '}\rightarrow 0$.
\end{description}

Substituting the expressions for $P_3(B)$ and $P_4(B)$ in Eq.
(\ref{Eq2}) we get \be
\bar{\gamma}_{12}^{(c)}(z)=(1+a_{1}^{'}{z}+a_{2}^{'}{z^{2}})^{-1}(1-d^{\
'}z^{-1})^{-1}\bar{\gamma}_{12}(z). \ee We now expand both sides in
powers of $z$:
\begin{eqnarray}
&& \cdots + \gamma_{12}^{(c)}(-1) z^{-1} + \gamma_{12}^{(c)}(0)+ \gamma_{12}^{(c)}(1) z + \cdots =
(1-a_{1}^{'}{z}+(a_1^2-a_{2}^{'}){z^{2}}+ \cdots ) \nonumber \\
&& \times (1+d^{\ '}z^{-1}+ d^{'2} z^{-2} + \cdots )
(\cdots + \gamma_{12}(-1) z^{-1} + \gamma_{12}(0)+ \gamma_{12}(1) z +  \cdots).
\end{eqnarray}
Collecting terms proportional to $z^{-1}, z^{0}, z^{1}$ etc., we
obtain the following expressions for the cross covariances at lag -1, 0 and 1:
\begin{eqnarray}
\gamma_{12}^{(c)}(-1)&=&
d^{\ '}(1-a_{1}{'}d^{\ '}+\ldots)(\gamma_{12}(0)+d^{\ '}\gamma_{12}(1)+\ldots),\\
\gamma_{12}^{(c)}(0)&=&(1-a_{1}^{'}d^{\ '}+\ldots)(\gamma_{12}(0)+d^{\ '}\gamma_{12}(1)+\ldots),\\
\gamma_{12}^{(c)}(1)&=&\gamma_{12}(1)-a_{1}^{'}\gamma_{12}(0)-a_{1}^{'}d^{\ '}\gamma_{12}(1)+\ldots \ .
\end{eqnarray}
We observe that $\gamma_{12}^{(c)}(k)$ for $k<0$ (and in particular,
$\gamma_{12}^{(c)}(-1)$) is no longer zero, implying that the
$X^{(c)}$ drives $Y^{(c)}$, thus giving rise to a spurious causal
direction. The spurious causality term $\gamma_{12}^{(c)}(-1)$ is
proportional to $d^{\ '}$.  This can be shown to be true for all the
other spurious terms $\gamma_{12}^{(c)}(k), \ k<-1$ as well. Hence
they all go to zero if $d^{\ '}\rightarrow 0$ (i.e. if $ {Y}$ has no
measurement noise).  This happens even if $a_{1}^{'}$ and
$a_{2}^{'}$ are non-zero (i.e. even if $ {X}$ measurement is
contaminated by noise). Hence we arrive at an important conclusion
that if $ {Y}$ is driving $ {X}$, only measurement noise in $ {Y}$
can cause spurious causality. If $ {Y}$ has no measurement noise, no
amount of measurement noise in $ {X}$ can lead to spurious
causality. Further, using the asymptotic properties of $d^{\ '}$
listed earlier, we can easily see that the magnitude of the spurious
causality increases as $d \rightarrow 1$ and as the ratio
$\sigma_{\eta}^2/{\sigma_{\eta'}^2} \rightarrow 0$.

The foregoing demonstrates that noise can lead to spurious causal
influences that are not part of the underlying processes. Here we
show that the true causality terms ($\gamma_{12}(k)$ for $k>0$) are
also modified by the presence of noise. They undergo a change even
if $d^{\ '}=0$. For example, $\gamma_{12}(1)$ is changed to
$\gamma_{12}(1)-a_{1}^{'}\gamma_{12}(0)$ even if $d^{\ '}=0$.
Therefore, it is quite possible that even a true causal direction
can be masked by added noise and the measurement noises in both time
series contribute to this suppression. As the ratios
${\sigma_{\xi}^{2}}/{\sigma_{\xi'}^{2}}$ and
$\sigma_{\eta}^{2}/\sigma_{\eta'}^{2}$ $\rightarrow\infty$,
$a_{1}^{'},a_{2}^{'},d^{\ '}$ all go to zero and
$\gamma_{12}^{(c)}\rightarrow \gamma_{12}$, as expected.

We make one final observation. If we replace $ {z}$ by $ {e^{i2 \pi
f}}$ (where $f$ is the frequency) in the covariance generating
function [cf. Eq. (\ref{covgenfn})] we obtain the cross spectrum.
Hence all the above results carry over to the spectral/frequency
domain.

To illustrate the above theoretical results, we estimate Granger
causality spectrum (in the frequency domain) for a bivariate AR
process numerically. First, we briefly summarize the theory behind
this computation \cite{Ding2}. The bivariate AR process given in Eq.
(\ref{bivariate_AR}) can be written as: \be \sum_{k=0}^{p} A(k)
Z(t-k) = E(t), \ee where $Z(t) = [X(t), Y(t)]^T $; $ E(t) = [E_1(t),
E_2(t)]^T $ and \be
 A(k) = \left (\begin{array}{cc}
-a_k & -b_k \\
-c_k & -d_k \end{array} \right ),
\ee
for $1 \leq k \leq p $.  $A(0)$ is the $2 \times 2$ identity matrix.  Here,
$E(t)$ is a temporally uncorrelated residual error with covariance
matrix $\Sigma$.  We obtain estimates of the coefficient matrices
$A(k)$ by solving the multivariate Yule-Walker equations
\cite{chatfield} using the Levinson-Wiggins-Robinson (LWR) algorithm \cite{morf_1978}.  From
$A(k)$ and $\Sigma$ we estimate the spectral
matrix $S(f)$ by the relation
\be
S(f) = H(f)\Sigma H^{*}(f),
\ee
where $H(f) = [\sum_{k=0}^{p} A(k) e^{-2\pi ikf}]^{-1} $ is the
transfer function of the system.

The Granger causality spectrum from $Y$ to $X$ is given by
\cite{Ding2,Geweke} (see also \cite{Hosoya}) \be I_{Y \rightarrow
X}(f) = - \ln \lbrack 1 - \frac{(\Sigma_{22} -
\frac{{\Sigma_{12}}^2}{\Sigma_{11}}   ){|H_{12}(f)|}^2 } {S_{11}(f)
} \rbrack . \ee Here, $\Sigma_{11}$, $\Sigma_{22}$ and $\Sigma_{12}$
are the elements of $\Sigma$ and $S_{11}(f)$ is the power spectrum
of $X$ at frequency $f$.  $H_{ij}(f)$ is the $\{ij\}^{th}$ element
of the transfer function matrix $H(f)$. Similarly, the Granger
causality spectrum from X to Y is defined by \be I_{X \rightarrow
Y}(f) =  - \ln \lbrack 1 - \frac{(\sum_{11} -
\frac{{\sum_{12}}^2}{\sum_{22}}   ){|H_{21}(f)|}^2 } {S_{22}(f) }
\rbrack, \ee and $S_{22}(f)$ is the power spectrum of $Y$ at
frequency $f$.

We now estimate the Granger causality spectrum for the specific
AR(1) process given in Eq. (\ref{Eq3}) where $Y$ drives $X$ and $X$
does not drive $Y$. The parameter values used are $a=0.4$, $b=0.6$,
$d=0.9$, $\sigma_{\xi}=0.2$ and $\sigma_{\eta}=1.0$.  We obtain two
time series $X$ and $Y$ by numerically simulating the VAR model and
then adding Gaussian measurement noise with $\sigma_{\xi'}=0.2$ and
$\sigma_{\eta'}=2.5$. For concreteness we assume that each time unit
corresponds to 5 ms. In other words, the sampling rate is 200 Hz,
and thus the Nyquist frequency is 100 Hz. The dataset consists of
one hundred realizations, each of length 250 ms (50 points). These
100 realizations are used to obtain expected values of the
covariance matrices in the LWR and KEM algorithms (see next
section). The Granger causality spectra $I_{X \rightarrow Y}(f)$ and
$I_{Y \rightarrow X}(f)$ are plotted in Figure 1. The solid lines
represents the true causality spectra while the dashed lines
represent the noisy causality spectra.

Similarly, we also simulated the following bivariate AR(2) process:
\begin{eqnarray}
 {X(t)}&=& a  {X(t - 1)} + b  {Y (t - 1)} +  E_1(t), \nonumber \\
 {Y(t)}&=& d_1 {Y(t-1)} + d_2 {Y(t-2)} + E_2(t).
\label{AR2process}\end{eqnarray} The values of the parameters $a$
and $b$ used were the same as in the previous AR(1) process example
(Eq. \ref{Eq3}) except for the values of the new parameters $d_1$
and $d_2$ which were chosen to be 0.4 and 0.5 respectively.  We
again obtain two time series $X$ and $Y$ and then added Gaussian measurement
noise with $\sigma_{\xi'}=0.2$ and $\sigma_{\eta'}=2.5$ to $X$ and
$Y$ respectively.  The Granger causality spectra $I_{X \rightarrow
Y}(f)$ and $I_{Y \rightarrow X}(f)$ are plotted in Figure 2.  As
before, the solid lines and dashed lines represent the true
causality spectra and noisy causality spectra, respectively.

We observe that the measurement
noise has a dramatic effect in both of these cases: It completely
reverses the true causal directions. For the noisy data, $X$ appears
to drive $Y$ and $Y$ does not appear to drive $X$.

The above theoretical and numerical results bring out clearly the
adverse effect that noise can have on correctly determining
directional influences. The same is also true for other quantities
like power spectrum and coherence. Therefore it is imperative that
the effect of noise be mitigated to the extent possible.

\section{The KEM Denoising Algorithm}

In the previous section we have seen that noisy data can lead to
grave misinterpretation of directional influences. We now provide a
practical solution to this problem by combining the Kalman smoother
with the Expectation-Maximization algorithm \cite{EM}. The detailed
algorithm is long and tedious. We outline the main logical steps
below.

Kalman filter \cite{Kalman} is a standard algorithm for denoising
noisy data. To apply this, we first need to recast a VAR process
with measurement noise in the so-called state-space form. This is
nothing but the difference equation analogue of converting a higher
order differential equation to a system of first order differential
equations. Once this is done, our VAR model takes on the following
form:
\begin{eqnarray}
\mathbf{x}_{t+1}&=&A\mathbf{x}_{t}+\mathbf{w}_{t+1},\\
\mathbf{y}_{t}&=&C\mathbf{x}_{t}+\mathbf{v}_{t}.
\end{eqnarray}
Here $\mathbf{x}_{t}$ is an  $M \times 1$ (``true'') state vector at
time $t$. $A$ is an $M \times M$ state matrix. $\mathbf{w}_{t}$ is a
zero mean Gaussian independent and identically distributed random
variable with covariance matrix $Q$. Bivariate AR(p) models can be
put in the form $\mathbf{x}_{t+1}=A\mathbf{x}_{t}+\mathbf{w}_{t+1}$
by defining $M = 2p$ auxiliary variables $x_{i,t}$. The $N \times 1$
vector $\mathbf{y}_{t}$ is the observed/measured value of $\mathbf
{x}_t$ in $N$ channels. $C$ is an $N \times M$ observation matrix
and is a fixed, known matrix for VAR models. Hence we will ignore
this in future discussions. The $N \times 1$ vector $\mathbf {v}_t$
is the measurement  noise which is zero mean, Gaussian, independent
and identically distributed with covariance matrix $R$.

Kalman filter, however, can not be directly applied to denoise
experimental or observed data since it assumes the knowledge of the
model describing the state space dynamics. In practice, such
knowledge is often not available. To get around this problem, we
apply the Kalman smoother in conjunction with the Expectation and
Maximization algorithm \cite{EM,Gahramani,Weinstein,Digalakis}.
Thus, this denoising algorithm will henceforth be called the KEM
algorithm. In this algorithm, one follows the standard procedure for
estimating state space parameters from data using the maximum
likelihood method. The appropriate likelihood function in our case
is the joint log likelihood log$P(\{\mathbf x\},\{\mathbf y\})$
where $\{\mathbf x\}$ denotes $\{\mathbf x_t\}$ (for all $t$) and
similarly for $\{\mathbf y\}$. In the usual maximum likelihood
method, $P$ would not depend on ${\mathbf x}$ and we would therefore
maximize the above quantity directly (conditioned on the observed
$\mathbf{y}_{t}$ values) and obtain the unknown state space
parameters.  But in our case, $P$ depends on $\mathbf{x}$ which is
also unknown. To get rid of $\mathbf{x}$, we take the expected value
of the log likelihood
$$O = E[\log P(\{\mathbf x\},\{\mathbf y\})\mid \{\mathbf y\}].$$
As usual, we have conditioned the expectation on the known
observations $\{\mathbf y\}$.

To compute $O$, it turns out we need the expectations of
$\mathbf{x}$ and $\mathbf{x} \mathbf{x}^T$ (where $T$ denotes the
transpose) conditioned on $\mathbf{y}$. These expectations are
obtained by applying the Kalman smoother on the noisy data. We use
the Kalman smoother and not the Kalman filter since we are utilizing
all the observations $\mathbf{y}$ instead of only the past
observations. This is the appropriate thing to do in our case since
we are performing an off-line analysis where all observations are
known. In other words, in Kalman smoother, we perform both a forward
pass and a backward pass on the data in order to make use of all
observations.

To apply the Kalman smoother, however, we still need the state space
model parameters (just as in the Kalman filter case). To circumvent
this problem, we start with initial estimates for these parameters
($A$, $Q$ and $R$) as follows. From the noisy data, using the LWR
algorithm, we obtain the VAR model coefficient matrices \cite{Ding}.
Then a standard transformation \cite{Kalman} is used to put these
matrices in the state space form giving the initial estimate for
$A$. The initial estimate of $Q$ is taken to be the identity matrix
following the standard procedure \cite{Kalman}. The initial estimate
of $R$ is taken to be half the covariance matrix at lag zero of the
noisy data. The approximate model order can be determined by
applying the AIC criterion \cite{akaike} in the LWR algorithm. This
step is admittedly rather ad hoc. Further studies to optimize the
above initial estimates and the VAR model order $p$ are currently
being carried out. Once we have initial estimates of the model
parameters, we can apply the Kalman smoother to obtain the various
conditional expectations and evaluate the expected log likelihood
$O$. This is called the expectation (E) step.

Next, we go to the maximization (M) step. Each of the parameters
$A,Q,R$ etc is re-estimated by maximizing $O$. Using these improved
estimates, we can apply the E step again followed by the M step.
This iterative process is continued till the value of log likelihood
function converges to a maximum. We could now directly use the VAR
parameters estimated from the KEM algorithm for further analysis as
is usually done. But here we prefer to use the following procedure
which was found to yield better performance. The final denoised data
(that is, the estimate of $\mathbf{x}$ obtained from the KEM algorithm)
is treated as the new experimental time series and subjected to
parametric spectral analysis from which Granger causality measures
can be derived. The Matlab code implementing this algorithm for our applications
is available from the authors upon request.

We have compared the denoising capabilities of the KEM algorithm with
two widely used algorithms, the higher-order Yule-Walker (HOY) method \cite{Chan}
and the
overdetermined higher-order Yule-Walker method \cite{Cadzow}. We find that the denoising
capabilities of the KEM algorithm is superior. Detailed results will be
presented elsewhere. In Figure 3, we explicitly show that KEM algorithm
performs better than the HOY method (see below).

The KEM algorithm is applied to denoise the data shown in Figures 1
and 2.  Figure 3 displays the same exact Granger causality spectra
(solid lines) as that in Figure 1 and the Granger causality spectra
(dashed lines) obtained from the denoised data using KEM algorithm.
Causality spectra obtained using HOY method is also shown (as dotted
lines). It is clear that the KEM method performs better. In Figure
4, the solid lines again represent the same exact Granger causality
as that in Figure 2 and the dashed lines represent the Granger
causality spectra obtained from the denoised data of a bivariate
AR(2) process.  We see that the correct causal directions are
recovered and that the denoised spectra are reasonably close to the
true causality spectra for both AR(1) and AR(2) process.  We stress
that these results are achieved without assuming any knowledge of
the VAR models [Eqs. \ref{Eq3} and \ref{AR2process}] that generated
the original time series data.

\section{Causal relations in a neural network model}

In this section, we analyze the effect of noise on time series
generated by a neural network model. We first demonstrate the effect
of measurement noise on causality directions and then the effect of
applying the KEM algorithm on the noisy data.

Our simulation model comprises two coupled cortical columns where
each column is made up of an excitatory and an inhibitory neuronal
population \cite{kaminski_2001}. The equations governing the dynamics of
the two columns are given by
\begin{eqnarray}\label{x1equ}
\ddot{x_{i}}+(a+b)\dot{x_{i}}+ab x_{i}  & = &
      -k_{ei}Q(y_{i}(t),Q_{m0}) + k_{ij}Q(x_{j}(t),Q_{m0}) +
             \xi_{x_{i}}(t), \\
\ddot{y_{i}}+(a+b)\dot{y_{i}}+aby_{i}  & = &
      k_{ie}Q(x_{i}(t),Q_{m0}) + \xi_{y_{i}}(t),
\end{eqnarray}
where $i \neq j = 1,2$. Here $x$ and $y$ represent local field
potentials (LFP) of the excitatory and inhibitory populations
respectively, $k_{ie} > 0$ gives the coupling gain from the
excitatory $(x)$ to the inhibitory $(y)$ population, and $k_{ei}>0$
is the strength of the reciprocal coupling. The neuronal populations
are coupled through a sigmoidal function $Q(x,Q_{m0})$ which
represents the pulse densities converted from $x$ with $Q_{m0}$ a
modulatory parameter. The function $Q(x,Q_{m0})$ is defined by \be
Q(x,Q_{m0}) = \left\{ \begin{array}{ll}
                Q_{m0}[1-e^{-(e^x - 1)/Q_{m0}} ] &
                        \mbox{if $x>-u_{0}$ } \\
                -1 & \mbox{if $x\leq -u_{0}$}
                \end{array} \right. ,
\ee
where $u_{0} = -\ln[1+\ln(1+\frac{1}{Q_{m0}})].$
The coupling strength $k_{ij}$ is the gain from the excitatory
population of column $j$ to the excitatory population of column
$i$, with $k_{ij}=0$ for $i=j.$  The terms $\xi(t)$ represent
independent Gaussian white noise inputs given to each neuronal
population.

The parameter values used were: $ a=0.22/$ms, $b=0.72/$ms,
$k_{ie}=0.1, k_{ei}=0.4, k_{12}=0, k_{21}=0.25$ and $Q_{m0}=5$.  The
standard deviation for the Gaussian white noise was chosen as 0.2.
Assuming a sampling rate of 200Hz, two hundred realizations of the
signals were generated, each of length 30 s (6,000 points).

We now restrict our attention to the variables $x_1(t)$ and
$x_2(t)$. Measurement noises (Gaussian white noises with standard
deviations 2.0 and 3.0 respectively) were added to these variables.
From the model it is clear that $x_1(t)$ should drive $x_2(t)$ since
$k_{12}=0$ while $k_{21}=0.25$. The results of applying Granger
causality analysis (using a VAR model of order 7) on these two variables is shown in Figure 5. The
solid lines represent the causality spectra for the noise-free data.
The dashed lines represents the causality spectra for the noisy
data. It is clear that the measurement noise has an effect on the
causal relations by significantly reducing the true causality
magnitude. In contrast to the example in Section 3, however, no
spurious causal direction is generated here, despite the fact that
both time series are contaminated by measurement noise. Next, we
applied the KEM algorithm to denoise the noisy data. When Granger
causality analysis is performed on the denoised data, we obtain
causality spectra that are closer to the true causality spectra (see
Figure 6). We note that the KEM algorithm is not able to completely
remove the noise as the denoised spectra are still quite different
from the true spectra.

To show that the denoised Granger spectrum is significantly
different from that of the noisy data we use the bootstrap approach
\cite{efron} to establish the significant difference between the two
peaks observed in Granger causality spectrum of Figures 5 and 6
(shown by dashed lines in these Figures). One thousand resamples of
noisy data and the denoised data were generated by randomly
selecting trials with replacement.  It should be noted that in any
selected trial, the entire multichannel data is taken as it is
thus preserving the auto and cross correlation structures. Thus,
we employ a version of block bootstrap method \cite{efron}. The peak values of Granger
causality were computed for each resample using both noisy data and
denoised data.  Let us denote these peak values by the random
variables $Z_1$ and $Z_2$ respectively. The two population Student
t-test was performed to determine whether the means of $Z_1$ and
$Z_2$ are different at a statistically significant level.

The null hypothesis was that the means of the two populations $Z_1$
and $Z_2$ are equal.  The $t$ value was found to be very
large: $4.6446*10^3$ and corresponds to a two-tailed $p$ value less
than 0.0001. Thus the null hypothesis that the two groups do not
differ in mean is rejected. This establishes the fact that the peak
of the Granger causality spectrum of the denoised data is
significantly higher than that of the noisy data.  Figure 7 shows
the plot of Granger causality for the direction $x_1 \rightarrow
x_2$ along with 95\% confidence intervals. The 95\% confidence intervals
are calculated as $I_{x_1 \rightarrow x_2}(f) \pm 1.96 \sigma_B$ (for each
frequency $f$) where
$\sigma_B$ is the sample standard deviation of the $1000$ bootstrap
replications of $I_{x_1 \rightarrow x_2}(f)$.

\section{Conclusions}

Our contributions in this paper are two fold. First, we demonstrate
that measurement noise can significantly impact Granger causality
analysis. Based on analytical expressions linking noise strengths
and the VAR model parameters, it was shown that spurious causality
can arise and that true causality can be suppressed due to noise
contamination. Numerical simulations were performed to illustrate
the theoretical results. Second, a practical solution to the
measurement noise problem, called the KEM algorithm, was outlined,
which combines the Kalman filter theory with the Expectation and
Maximization (EM) algorithm. It was shown that the application of
this algorithm to denoise the noisy data can significantly mitigate
the deleterious effects of measurement noise on Granger causality
estimation. It is worth noting that, despite the fact that the
adverse effect of measurement noise on Granger causality has been
known since 1978 \cite{Newbold}, mitigation of such effect has
received little attention. The KEM algorithm described in this paper
is our attempt at addressing this shortcoming.

\section*{Acknowledgements}

This work was supported by NIH grant MH071620.
GR was supported in part by grants from DRDO and UGC (under
DSA-SAP Phase IV). GR is also a Honorary Faculty Member of the
Jawaharlal Nehru Centre for Advanced Scientific Research, Bangalore.

\appendix
\section*{Appendix}

In this appendix, we derive the expressions for $P_3(B)$ and
$P_4(B)$ given in Eq. (\ref{P3P4}). We first determine $P_4(B)$.
When a zero mean white noise process $\eta'(t)$ with variance
$\sigma_{\eta'}^{2}$ is added to $Y(t)$ we get \be {Y^{(c)}(t)}=
{Y(t)+\eta'(t)}. \ee Applying $(1-dB)$ on both sides of the above
equation we get
\begin{eqnarray}
(1-dB) {Y^{(c)}(t)}& = &(1-dB) {Y(t)} + (1-dB) {\eta'(t)}\nonumber \\
 & = &  {\eta(t)}+ (1-dB) {\eta'(t)}.
\end{eqnarray}
We now determine a white noise process $ {\eta^{(c)}(t)}$ such
that
\be
{\eta(t)} + (1-dB) {\eta'(t)} = (1-d'B) {\eta^{(c)}(t)}.
\ee
We need to determine $d'$ and $ {\sigma_{\eta^{(c)}}^{2}}$.

Taking variances on both sides of the above equation we get
\be
{\sigma_{\eta}^{2}} + (1+d^{2})  {\sigma_{\eta'}^{2}} = (1 + d'^{2}) {\sigma_{\eta^{(c)}}^{2}}.
\ee
Taking autocovariance at lag 1 on both sides we obtain
\be
d  {\sigma_{\eta'}^{2}} = d'  {\sigma_{\eta^{(c)}}^{2}}.
\ee
Since $ {\eta^{(c)}}$ is a sum of $ {\eta}$ and $(1-dB) {\eta'}$,
we have ${\sigma_{\eta^{(c)}}^{2}} > { {\sigma_{\eta'}^{2}}}.$
This implies that $|d'| < |d|$.  Since stationarity of the AR
process requires $ 0 < |d| < 1$, we obtain the inequality $0 <
|d'| < |d'|<1.$ Further $d'$ has the same sign as $d$.

We have
\be
{\sigma_{\eta^{(c)}}^{2}}= \frac{d}{d'}
{\sigma_{\eta'}^{2}}.
\ee
Substituting in the variance equation we
get
\be
(1 + d~^{{'}{2}})\frac{d}{d'} {\sigma_{\eta'}^{2}} = (1 + d^{2})
{\sigma_{\eta'}^{2}}+ {\sigma_{\eta}^{2}},
\ee
that is,
\be
(\frac{1}{d'}+d')=(\frac{1}{d}+d)+\frac{1}{d}\frac{
{\sigma_{\eta}^{2}}}{ {\sigma_{\eta'}^{2}}}.
\ee
Let
\be
s\equiv(\frac{1}{d}+d)+\frac{1}{d}\frac{
{\sigma_{\eta}^{2}}}{ {\sigma_{\eta'}^{2}}}.$$ This gives
$$(\frac{1}{d'}+d') = s.
\ee
Hence
\be
d' = \frac{s{\pm{\sqrt{s^{2}-{4}}}}}{2}.
\ee

Note that $ |s| > 2 $ for any value of $d$,
 ${\sigma_{\eta}^{2}}$, and ${\sigma_{\eta'}^{2}}$.
Therefore $\sqrt{s^{2}-{4}}$ and hence $d{'}$ are well defined.
Further, since $|d{'}| < |d|$ if $d$  is positive,  $d' = ({s
- \sqrt{s^{2}-{4}}})/{2}$ is the only valid solution. If $d$  is
negative,  $d' = ({s +\sqrt{s^{2}-{4}}})/{2} $ is the only valid
solution.

Next, we derive the expression for $P_3(B)$. First, we first need
to rewrite $ {X(t)}$ as an univariate process i.e. we need to
determine $P_1(B)$:
\be
P_{1}(B) X(t)=\xi(t),
\ee
where $\xi(t)$ is a zero mean white noise process and
\be
 {X(t)}= a {X(t-1)}+b {Y(t-1)}+E_1(t).
\ee
Here $E_1(t)$ is a zero mean white noise process with variance $\sigma_{\epsilon}^2$.
We have already seen that
\be
(1-dB) {Y(t)}={\eta(t)}.
\ee
The equation for $X(t)$ can be written as
\be
(1-aB) {X(t)}= b {Y(t-1)}+E_1(t).
\ee
Substituting the expression for $Y(t-1)$ we obtain
\be
(1-aB) {X(t)}=b(1-dB)^{-1} {\eta(t-1)}+E_1(t).
\ee

We now find a white noise
process $ {\xi(t)}$ with variance $ {\sigma_{\xi}^{2}}$ such that
\be
b(1-dB)^{-1} {\eta(t-1)}+E_1(t)= (1-rB)^{-1} {\xi(t)}.
\ee
To determine $ {r}$ and $ {\sigma_{\xi}^{2}}$, we take variance
and autocovariance at lag 1 on both sides.  Taking variance we
obtain
\be
\frac{b^{2} {\sigma_{\eta}^{2}}}{(1-d^{2})}+ {\sigma_{\epsilon}^{2}}=\frac{ {\sigma_{\xi}^{2}}}{(1-r^{2})}.
\ee
Taking autocovariance at lag 1 and assuming that $\sigma_{\epsilon
\eta}$ (the cross-covariance between $E_1$ and $\xi$) is zero for simplicity, we get
\be
\frac{b^{2} \sigma_{\eta}^{2}d}{(1-d^{2})}=\frac{ {\sigma_{\xi}^{2}}}{(1-r^{2})},
\ee
which can be written as
\be \frac{ {\sigma_{\xi}^{2}}}{(1-r^{2})}=\frac{b^{2} {\sigma_{\eta}^{2}}{d}}{(1-d^{2})r}.
\ee
Substituting in the variance equation we obtain
\be
\frac{b^{2} {\sigma_{\eta}^{2}}}{(1-d^{2})}+
{\sigma_{\epsilon}^{2}}=\frac{b^{2}
{\sigma_{\eta}^{2}}}{(1-d^{2})}\frac{d}{r}.
\ee
Thus
\be
r=\frac{b^{2}d {\sigma_{\eta}^{2}}}{b^{2} {\sigma_{\eta}^{2}}+(1-d^{2}) {\sigma_1^{2}}}.
\ee
If $b = 0$, we get $r = 0$ and $ {\sigma_{\xi}^{2}}=
{\sigma_1^{2}}$ as expected. Similarly if $d=0$, we get $r =0$ and
$ {\sigma_{\xi}^{2}}= {\sigma_1^{2}}+b^{2} {\sigma_{\eta}^{2}}$ as
expected. Once $r$ is known, $ {\sigma_{\xi}^{2}}$ is given by
\be
{\sigma_{\xi}^{2}}=(1-r^{2})\left[\frac{b^{2} {\sigma_{\eta}^{2}}}{(1-d^{2})}+ {\sigma_1^{2}}\right].
\ee

We finally have
\be
(1-aB) {X(t)}= (1-rB)^{-1} {\xi(t)}.
\ee
That is,
\be
P_1(B) X(t)=\xi(t), \ \ \ P_{1}{(B)}=(1-rB)(1-aB).
\ee

Consider a white noise process $\xi'(t)$ (which is
uncorrelated with $X(t)$) and has variance $\sigma_{\xi'}^{2}$.
This is added to $X(t)$ to obtain the noisy process $X^{(c)}(t)$:
\be X^{(c)}(t)= X(t)+\xi'{(t)}.
\ee
Applying $P_{1}(B)$ on both sides of the above equation,
\be
P_{1}(B)X^{(c)}(t)=\xi(t)+P_{1}{(B)}\xi'{(t)}.
\ee
We need to find a zero mean white noise process $\xi^{(c)}{(t)}$
with variance $\sigma_{\xi^{(c)}}^{2}$ such that
\be
\xi(t)+P_{1}{(B)}\xi'{(t)}=P_{3}{(B)}\xi^{(c)}{(t)}.
\ee
Let
\be
P_{3}(B)= 1 + a_{1}^{'}B+a_{2}^{'}B^{2}.
\ee
We have
\be
\xi(t)+(1-(a+r)B+arB^{2})\xi'(t)= [1 + a_{1}^{'}B+a_{2}^{'}B^{2}]\xi^{(c)}(t).
\ee

Taking variances on both sides we get
\be\label{eq.68}
\sigma_{\xi}^{2}+(1+(a+r)^{2}+a^{2}r^{2})\sigma_{\xi'}^{2}=[1+a_{1}'^{{\ 2}}+a_{2}'^{{\ 2}}]\sigma_{\xi^{(c)}}^{2}.
\ee
Taking autocovariance at lag 1 on both sides we obtain
\be
-(a+r)\sigma_{\xi'}^{2}-ar(a+r)\sigma_{\xi'}^{2}=a_{1}^{'}\sigma_{\xi^{(c)}}^{2}+a_{1}^{'}a_{2}^{'}\sigma_{\xi^{(c)}}^{2}.
\ee
This can be rewritten as
\be\label{eq.70}
-(a+r)(1+ar)\sigma_{\xi'}^{2}=a_{1}^{'}(1+a_{2}^{'})\sigma_{\xi^{(c)}}^{2}.
\ee
Taking autocovariance at lag 2 on both sides
\be
ar\sigma_{\xi'}^{2}=a_{2}^{'}\sigma_{\xi^{(c)}}^{2},
\ee
which gives
\be
\sigma_{\xi^{(c)}}^{2}=\frac{ar}{a_{2}^{'}}\sigma_{\xi'}^{2}.
\ee
Since $\sigma_{\xi^{(c)}}^{2}>\sigma_{\xi'}^{2}$, we see that
$|a_{2}^{'}|<|ar|$ and $a_{2}^{\, '}$ has the same sign as $ar$.

Substituting the last equation in Eqs. (\ref{eq.70}) and (\ref{eq.68}) we
obtain
\be
-(a+r)(1+ar)\sigma_{\xi'}^{2}=a_{1}^{'}(1+a_{2}^{'})\frac{ar}{a_{2}^{'}}\sigma_{\xi'}^{2},
\ee
and
\be
\sigma_{\xi}^{2}+[1+(a+r)^{2}+a^{2}r^{2}]\sigma_{\xi'}^{2}=[1+a_{1}'^{{2}}+a_{2}'^{{2}}]\frac{ar}{a_{2}^{'}}\sigma_{\xi'}^{2}.
\ee
Thus we get
\be
\frac{a_{1}^{'}(1+{a_{2}^{'})}}{a_{2}^{'}}=-\frac{(a+r)(1+ar)}{ar},
\ee
and
\be
\frac{(1+a_{1}'^{{2}}+a_{2}'^{{2}})}{a_{2}'}=\frac{[1+(a+r)^{2}+a^{2}r^{2}]}{a_r}+\frac{1}{ar}\frac{ {\sigma_{\xi}^{2}}}{ {\sigma_{\xi'}^{2}}}.
\ee
We can solve these two equations for $a_1^{'}$ and $a_{2}^{'}$.
There will be multiple solutions.  We choose that solution for
which $|a_{2}^{'}|<|ar|$.  Further the solution has to be such
that the roots of $1+a_{1}^{'}B+a_{2}^{'}B^{2}=0$ lie outside the
unit circle. The last condition is required for the invertibility
of the MA process $(1 + a_{1}^{'}B+a_{2}^{'}B^{2})\xi^{(c)}(t)$.
The expressions for $a_1^{'}$ and $a_{2}^{'}$ obtained by solving
the above equations are very long and therefore we do not list
them here. However, we can easily obtain the asymptotic behaviour
of these solutions as follows.

For our bivariate AR(1)process to be stable, we require that the
roots of
\be \det[\lambda I-A(1)]=0
\ee
lie within the unit circle i.e., the eigenvalues of A(1) should
have absolute value less than 1.  In our case
\begin{displaymath}
 {A(1)}={\left(\begin{array}{cc} a&b\\
0&d\end{array}\right)},
\end{displaymath}
which is an upper triangular matrix.  Hence eigenvalues are $a$
and $d$.  Therefore, for stability we require that $ |a|<1$ and
$|d|<1$.

As already derived, we have
\be
r=d{\left(\frac{b^{2} {\sigma_\eta^{2}}}{b^{2}
{\sigma_\eta^{2}}+(1-d^{2}) {\sigma_\xi^{2}}}\right)}.
\ee
Since $|d|<1$, the term within brackets is always positive and
less than 1.  It becomes zero only when $b = 0$.  Hence $|r|<|d|$
and $r$ has same sign as $d$. As $|d|\rightarrow1$,
$|r|\rightarrow1$. As $|d|\rightarrow 0$ or $|b|\rightarrow 0$, we
see that $|r|\rightarrow 0$.

We have already seen that $|a_{2}^{'}|<|ar|$.  Since $|r|<|d|$, we
obtain further results that $|a_{2}^{'}|<|a||d|$ and $a_{2}^{'}$
has same sign as $ad$. Since $|a|, |d|<1$, we get
\[0<|a_{2}^{\, '}|<|a||d|<1.\]
As $|a|, |d|\rightarrow1$, $|a_{2}^{'}| ~\rm{also}~ \rightarrow1$.
As $a\rightarrow1$, $d\rightarrow1$ and  the ratio $
{\sigma_{\xi}^{2}}/ {\sigma_{\xi'}^{2}}\rightarrow0$, we have
$$a_{1}^{\, '}\rightarrow {-2}; a_{1}^{'}\rightarrow 1.$$
As the variance ratio $\rightarrow \infty$
$$a_{1}^{'}\rightarrow 0;a_{2}^{\, '}\rightarrow 0,$$ as expected. The
parameter $a_{1}^{\, '}$ is hardly affected by the value of the
parameter $b$. On the other hand, $a_{2}^{\, '}\rightarrow0$ as
$b\rightarrow0$ and saturates rapidly for $b>0.5$.



\begin{figure}[htbp]
\includegraphics[width=0.65\textwidth]{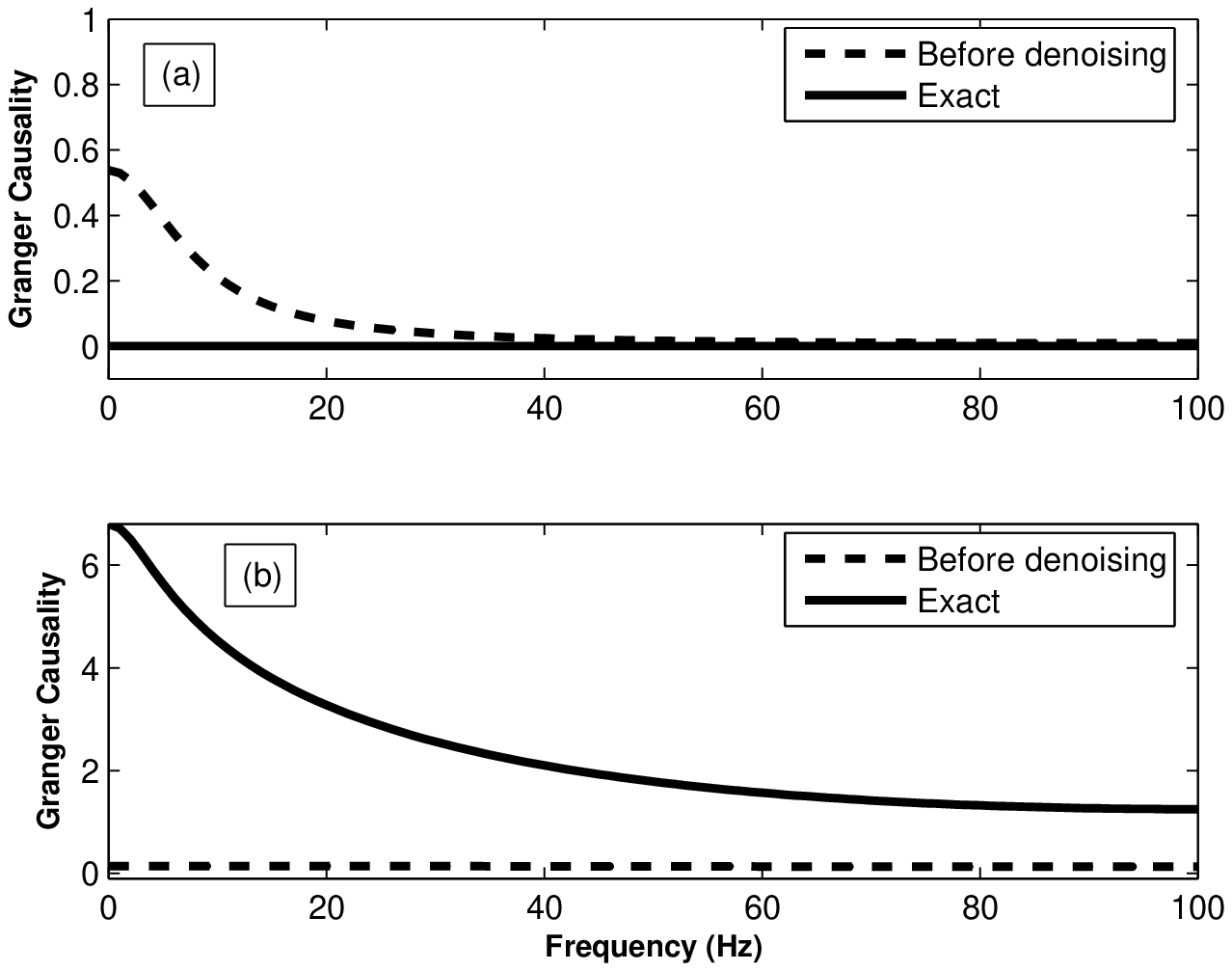}
\caption{Granger causality spectra for a bivariate AR(1)
process (a) Causality of $X \rightarrow Y$  (b) Causality of $Y
\rightarrow X$.  The solid lines represent true causality spectra
and the dashed lines represent spectra from noisy data.}
\label{Figure1}
\end{figure}

\begin{figure}[htbp]
\includegraphics[width=0.65\textwidth]{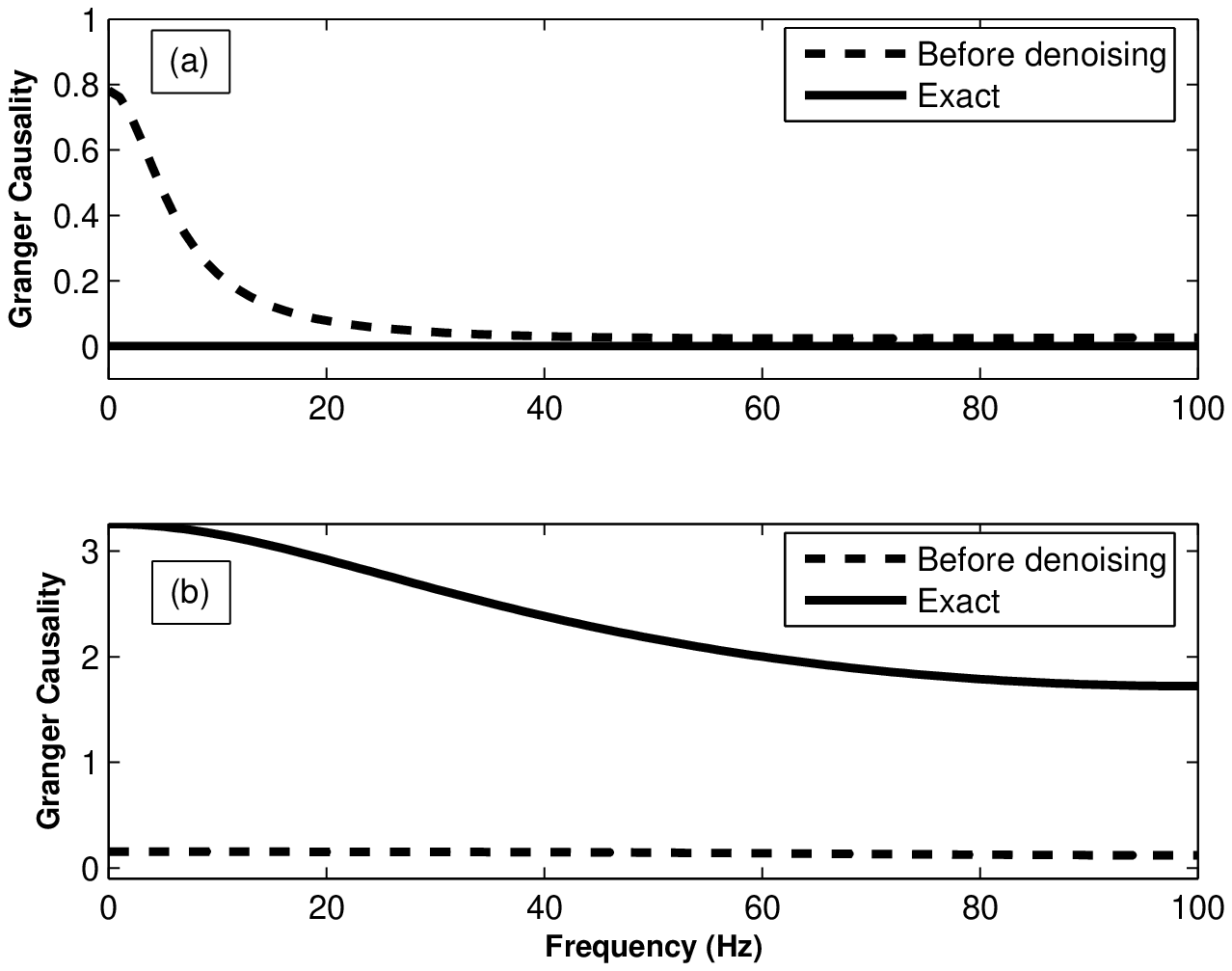}
\caption{Granger causality spectra for a bivariate AR(2) process
(a) Causality of $X \rightarrow Y$  (b) Causality of $Y \rightarrow
X$.  The solid lines represent true causality spectra and the dashed
lines represent spectra from noisy data.}
\label{Figure2}
\end{figure}

\begin{figure}[htbp]
\includegraphics[width=0.65\textwidth]{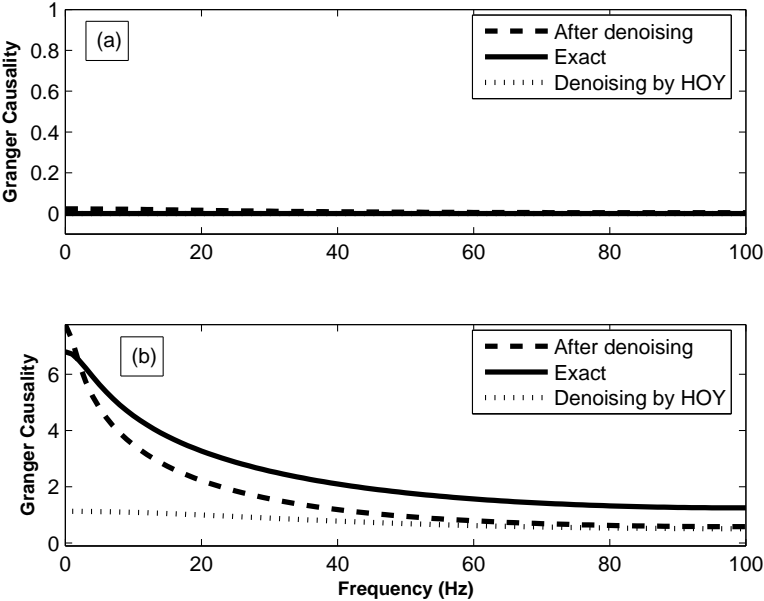}
\caption{Granger causality spectra for the bivariate AR(1) process in Fig 1.
(a) Causality of $X \rightarrow Y$  (b) Causality of $Y \rightarrow
X$.  The solid lines represent true causality spectra and the dashed
lines represent spectra obtained from the denoised data using the
KEM algorithm. The dotted lines represent spectra obtained using HOY
algorithm.}
\label{Figure3}
\end{figure}

\begin{figure}[htbp]
\includegraphics[width=0.65\textwidth]{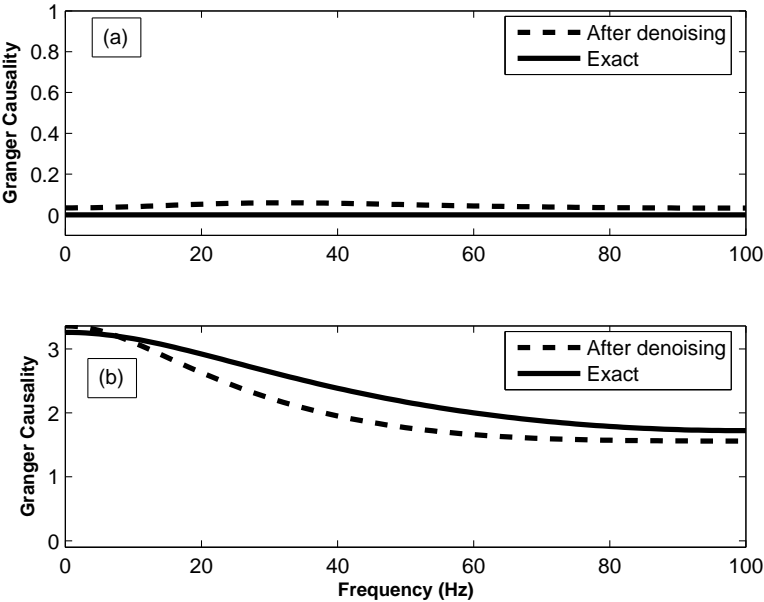}
\caption{Granger causality spectra for the bivariate AR(2) process in Fig 2.
(a) Causality of $X \rightarrow Y$  (b) Causality of $Y \rightarrow
X$.  The solid lines represent true causality spectra and the dashed
lines represent spectra obtained from the denoised data using the
KEM algorithm.}
\label{Figure4}
\end{figure}

\begin{figure}[htbp]
\includegraphics[width=0.65\textwidth]{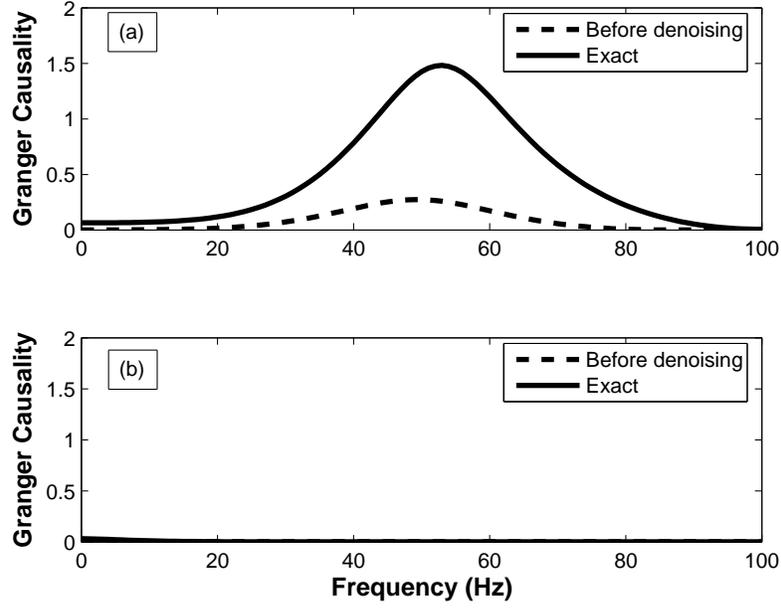}
\caption{Granger causality spectra for noisy data from a neural network model
(a) Causality of $x_1 \rightarrow x_2$  (b) Causality of $x_2
\rightarrow x_1$.  The solid lines represent true causality spectra
(noise-free data) and the dashed lines represent spectra from noisy
data.}
\label{Figure5}
\end{figure}

\begin{figure}[htbp]
\includegraphics[width=0.65\textwidth]{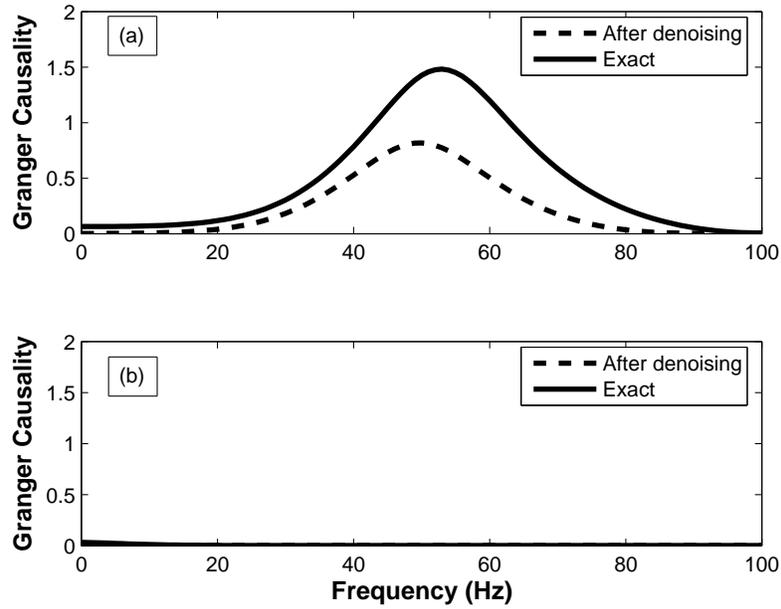}
\caption{Granger causality spectra of the neural network model
(a) Causality of $x_1 \rightarrow x_2$  (b) Causality of $x_2
\rightarrow x_1$.  The solid lines represent true causality spectra
(noise-free data) and the dashed lines represent spectra obtained
from denoised data using the KEM algorithm.}
\label{Figure6}
\end{figure}

\begin{figure}[htbp]
\includegraphics[width=0.65\textwidth]{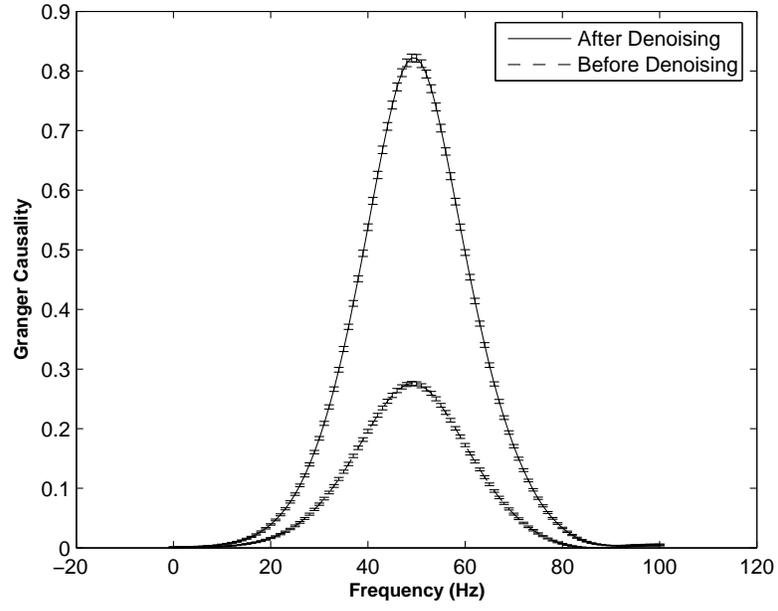}
\caption{Granger causality spectra of the neural network
model for the direction $x_1 \rightarrow x_2$.  The solid line
represents the Granger causality for denoised data, while the dashed
line represents the Granger causality for noisy data. 95\%
confidence intervals are also given..}
\label{Figure7}
\end{figure}

\end{document}